\documentclass{aastex}
\usepackage{emulateapj5,apjfonts,rotating}

\newcommand{\pcmsq}{\mbox{cm$^{-2}$}}
\newcommand{\cmsq}{\mbox{cm$^{2}$}}
\newcommand{\ergsec}{\mbox{erg s$^{-1}$}}
\newcommand{\ergcms}{\mbox{erg cm$^{-2}$ s$^{-1}$}}

\newcommand{\Lx}{\mbox{$L_{\rm x}$}}

\newcommand{\uJy}{\mbox{$\mu$Jy}}

\newcommand{\nh}{\mbox{$N_{\rm H}$}}
\newcommand{\chandra}{\textit{Chandra}}
\newcommand{\cxo}{\textit{Chandra X-ray Observatory}}

\newcommand{\rosat}{\textit{ROSAT}}
\newcommand{\xspec}{\textsc{XSPEC}}
\newcommand{\tool}{\it}
\newcommand{\xsoft}{\mbox{X$_{\rm soft}$}}
\newcommand{\xmed}{\mbox{X$_{\rm med}$}}
\newcommand{\xhard}{\mbox{X$_{\rm hard}$}}
\newcommand{\til}{$\sim$}
 
\newcommand{\hst}{\textit{HST}}
\newcommand{\hubble}{\textit{Hubble Space Telescope}}
\newcommand{\err}[2]{\small \ensuremath{^{+#1}_{-#2}}}
\newcommand{\ee}[2]{\ensuremath{#1\times 10^{#2}}}

\shorttitle{Faint X-ray Sources in NGC 6752}
\shortauthors{Pooley et al.}
\slugcomment{{\sc Accepted by the} {\it Astrophysical Journal}}

\begin{document}

\submitted{Accepted by the {\it Astrophysical Journal}}

\title{Optical Identification of Multiple Faint X-ray Sources in the
Globular Cluster NGC 6752: Evidence for Numerous Cataclysmic
Variables$^1$}\addtocounter{footnote}{1}\footnotetext{Based on observations with the NASA/ESA {\it Hubble
Space Telescope}, obtained at the Space Telescope Science Institute,
which is operated by AURA, Inc.\ under NASA contract NAS5-26555.}

\author{David Pooley\altaffilmark{2},
Walter H.\ G.\ Lewin\altaffilmark{2}, 
Lee Homer\altaffilmark{3},
Frank Verbunt\altaffilmark{4},
Scott F.\ Anderson\altaffilmark{3},
Bryan M.\ Gaensler\altaffilmark{2,5,6},
Bruce Margon\altaffilmark{7},
Jon M.\ Miller\altaffilmark{2}, 
Derek W.\ Fox\altaffilmark{8}, 
Victoria M.\ Kaspi\altaffilmark{9,2},
Michiel~van~der~Klis\altaffilmark{10}}

\altaffiltext{2}{Center for Space Research and Department of Physics,
Massachusetts Institute of Technology, Cambridge, MA  02139-4307;
davep@space.mit.edu, lewin@space.mit.edu, jmm@space.mit.edu} 
\altaffiltext{3}{Astronomy Department, Box 351580, University of
Washington, Seattle, WA 98195; homer@astro.washington.edu,
anderson@astro.washington.edu}
\altaffiltext{4}{Astronomical Institute, Utrecht University, PO Box
80000, 3508 TA Utrecht, The Netherlands; F.W.M.Verbunt@astro.uu.nl}
\altaffiltext{5}{Current address: Harvard-Smithsonian Center for
Astrophysics, 60 Garden Street, Cambridge, MA  02138;
bgaensler@cfa.harvard.edu}
\altaffiltext{6}{Hubble Fellow}
\altaffiltext{7}{Space Telescope Science Institute, 3700 San Martin
Drive, Baltimore, MD 21218; margon@stsci.edu}
\altaffiltext{8}{Astronomy Department, California Institute of
Technology, Mail Code 105-24, Pasadena, CA 91125;
derekfox@astro.caltech.edu}
\altaffiltext{9}{Department of Physics, Rutherford Physics Building, McGill University, 3600
University Street, Montreal, QC H3A 2T8, Canada; vkaspi@physics.mcgill.ca} 
\altaffiltext{10}{Astronomical Institute ``Anton Pannekoek,''
University of Amsterdam and Center for High-Energy Astrophysics,
Kruislaan 403, 1098 SJ Amsterdam, The Netherlands; michiel@astro.uva.nl}

\begin{abstract}
We report on the \cxo\ ACIS-S3 imaging observation of the globular cluster
NGC~6752. We detect 6 X-ray sources within the 10\farcs5 core radius
and 13 more within the 115\arcsec\ half-mass radius down to a limiting
luminosity of $\Lx\approx10^{30}$~\ergsec\ for cluster sources.  We
reanalyze archival data from the \hubble\ and the Australia Telescope
Compact Array and make 12 optical identifications and one radio
identification.  Based on X-ray and optical properties of the
identifications, we find 10 likely cataclysmic variables (CVs), 1--3
likely RS~CVn or BY~Dra systems, and 1 or 2 possible background
objects.  Of the 7 sources for which no optical identifications were
made, we expect that $\sim$2--4 are background objects and that the
rest are either CVs or some or all of the five millisecond pulsars whose
radio positions are not yet accurately known.  These and other
\chandra\ results on globular clusters indicate that the dozens of CVs
per cluster expected by theoretical arguments are being found.
The findings to date also suggest that the ratio of CVs to other types
of X-ray sources is remarkably similar in clusters of very different
structural parameters.

\end{abstract}


\section{Introduction}
Globular clusters are very efficient catalysts in forming unusual
objects, such as low-mass X-ray binaries (LMXBs), cataclysmic
variables (CVs), millisecond pulsars (MSPs), and blue stragglers, with
formation rates per unit mass exceeding those in the galactic disk by
orders of magnitude.  The high stellar densities in globular clusters
trigger various dynamical interactions: exchanges in encounters with
binaries, direct collisions, destruction of binaries, and perhaps
tidal capture.  Many details of these processes are not well
understood, primarily because of the complex feedback between stellar
evolution and stellar dynamics.  The rate of these competing
processes depends strongly on the cluster's physical properties.  For
example, binary formation by tidal capture generally requires higher
densities than do exchange reactions involving primordial binaries
\citep{hu92}.  In the outermost regions of even the densest clusters,
the currently observed binaries will be mainly primordial ones.  In
contrast, the LMXBs and the MSPs that consequently evolve from the LMXBs are
formed (almost) exclusively via stellar encounters.  \citet{dh98} have
shown how the large number of binary radio pulsars that are present in
47~Tuc \citep{cam00,fr01} may have been formed from an initial
population of some $10^{6}$ primordial binaries and $\sim$10$^{4}$
neutron stars.

To date, there are thirteen known bright ($\Lx \ga 10^{36}$~\ergsec)
X-ray sources in 12 globular clusters.  The X-ray spectra and
luminosities indicate that these are LMXBs; orbital periods from 11
minutes to 17.1 hours have been determined for five sources. In 11
sources, type I X-ray bursts have been detected (in 't Zand et al.\
1999 and references therein), which are well explained as
thermonuclear runaways on a neutron star surface (for a review see
Lewin, van Paradijs, \& Taam 1993).  Optical counterparts have been
found for six bright sources \citep[and references
therein]{deutsch98,heinke01,homer01}.

With the \textit{Einstein} satellite, seven faint ($\Lx \la
10^{35}$~\ergsec) sources were detected in the cores of as many
clusters \citep{hg83a,hg83b}.  With \rosat, the number of faint core
sources (within two core radii of the centers) expanded to 40,
including multiple sources in 47~Tuc, $\omega$~Cen, NGC~6397, NGC~6752
and others; an additional 17 faint sources were found farther out in
the clusters (more than two core radii from the centers).  Whereas
virtually all of the core sources are related to the globular
clusters, some of the sources outside the cores may be in the background or
foreground \citep[and references therein]{vb01}.  Due to
crowding and the limited accuracy of even the \rosat\ positions
(varying from 2--5\arcsec, depending on whether a secure optical
identification allows accurate determination of the bore sight
correction), optical identifications of faint X-ray sources in
globular cluster cores remained tentative before the \chandra\ era,
and some suggestions have been disproved with more accurate X-ray
positions, as in 47~Tuc \citep{vh98} and M~92 \citep{gef98,vb01}.  The
only secure optical identification was that of a dwarf nova with a
faint X-ray source 12 core radii from the center of NGC~5904
\citep{hak97,mar81}. The radio pulsar in M~28 is another securely
identified faint X-ray source \citep{ly87,sai97}.  Some
identifications, like those in M~13, remain in doubt \citep{vb01}.

Progress in identifying the nature of the various faint X-ray sources
has been very slow during the past two decades. However, this is now
rapidly changing.  With \chandra's high sensitivity and unprecedented
spatial resolution of $\sim$1\arcsec, a revolution is underway in our
understanding of the low-luminosity globular cluster X-ray sources and
their association with the binary populations of globular clusters.
To date, some results have already been reported for a 74~ks
observation of 47~Tuc \citep{gr01a}, a 70~ksec observation of
$\omega$~Cen \citep{rut01}, a 49~ks observation of NGC~6397
\citep{gr01b}, and preliminary results have been reported for our own
observations of NGC~6752, NGC~6440 and NGC~6121 \citep{po01}.  The
\chandra\ observations have confirmed previously suggested optical
counterparts for faint X-ray sources in the cores of these clusters
and made numerous additional identifications.  We report here in more
detail on the results of a 30~ks observation of NGC~6752. In the
discussion section we will compare our results with those reported
earlier.

NGC~6752 is at a distance of $4.1\pm 0.2$~kpc \citep{renz96}.  The
optically derived center is at (J2000) $19^{\rm h} 10^{\rm m} 51\fs8,
-59\degr 58\arcmin 55\arcsec$ \citep{har96}.  Its moderate optical
reddening $E_{B-V}=0.04$ may be converted to a nominal X-ray absorption
column characterized by $\nh=\ee{2.2}{20}$~\pcmsq\ using the relation
found by \citet{pred95}.  Its core radius $r_{\rm c}$ is 10\farcs5,
and its half-mass radius $r_{\rm h}$ is 115\arcsec\ \citep{tra93}.  We
use these values throughout the paper.

Faint X-ray sources in NGC~6752 were first found by \citet{gr93} on
the basis of a 31.3~ks \rosat\ HRI exposure, which revealed a double
source in the core and two sources outside the core but within the
half-mass radius. These sources were also detected with the \rosat\
PSPC \citep{johnst94}.  \citet{vj00} co-added three \rosat\ HRI
observations for a total exposure of 72~ks and identified an X-ray
source with the non-cluster member star TYC\,9071~228~1, allowing an accurate solution of the
bore sight correction and therewith an accurate
($\sigma\simeq2\arcsec$) absolute positioning of the X-ray frame. They
also resolved the central source into four sources and argued on the
basis of the source density in the whole \rosat\ image that the other
sources within the half-mass radius were possibly related to the
cluster.  The best position for the northernmost \rosat\ source in the
core is marginally compatible with the previously published position
of the southern of two H\,$\alpha$ emission objects and photometric
variables with periods of 5.1~hr and 3.7~hr discovered with \hubble\
(\hst) observations by \citet{ba96}.  

Our \chandra\ X-ray observations are described and analysed in
\S\ref{sect:xrayobs}, our radio observations in
\S\ref{sect:radio}, and our analysis of the \hst\ observations in
\S\ref{sect:optid}. The results are described in
\S\ref{sect:res} and discussed in \S\ref{sect:dis}.

\section{X-ray Observations} \label{sect:xrayobs}
NGC~6752 was observed with the \cxo\ \citep{weiss96} for $\sim$30~ks
on 2000~May~15.  The observation was made with the Advanced CCD
Imaging Spectrometer (ACIS) with the telescope aimpoint on the
back-side illuminated S3 chip, which offers increased sensitivity to
low-energy X-rays compared to the front-side illuminated chips.  The
entire region inside the cluster's half-mass radius fit on the
8\farcm3 square S3 chip.  The data were taken in timed-exposure mode
with the standard integration time of 3.24~sec per frame and
telemetered to the ground in faint mode, in which a $3\times3$ pixel
island is recorded for each event.

\subsection{X-ray Data Reduction}
Data reduction was performed using the CIAO~2.1 software provided by
the \chandra\ X-ray Center (\url{http://asc.harvard.edu}).  We used
the CALDB v2.6 calibration files (gain maps, quantum efficiency,
quantum efficiency uniformity, effective area).  Bad pixels were
excluded, as were intervals of background flaring ($\sim$1~ks).  The
effective exposure time for the observation after filtering for flares
and correcting for dead time was 28.7~ks.

Starting with the raw (level~1) event list, we processed the data
(using {\tool acis\_process\_events}) without including the pixel
randomization that is added during standard processing.\footnote{This
randomization has the effect of removing the artificial substructure
(Moir\'{e} pattern) that results as a byproduct of spacecraft dither.
Since all of our observations contained a substantial number of dither
cycles (one dither cycle has a period of $\sim$1000~sec), this
substructure is effectively washed out, and there is no need to blur
the image with pixel randomization.}  This method slightly improves
the point spread function (PSF).  We then applied good-time intervals
(both the ones supplied with the standard data products and our custom
ones which excluded the period of background flaring) and filtered the
data to include only events with {\it ASCA} grades of 0, 2, 3, 4, or 6
(this is the ``standard'' choice that generally optimizes the
signal-to-background ratio; see the \chandra\ Proposer's Observatory
Guide available from the website for more information).  We also
excluded software-flagged cosmic ray events.  We used this filtered
event list (level~2) for the subsequent analysis.

\begin{figure*}
\plotone{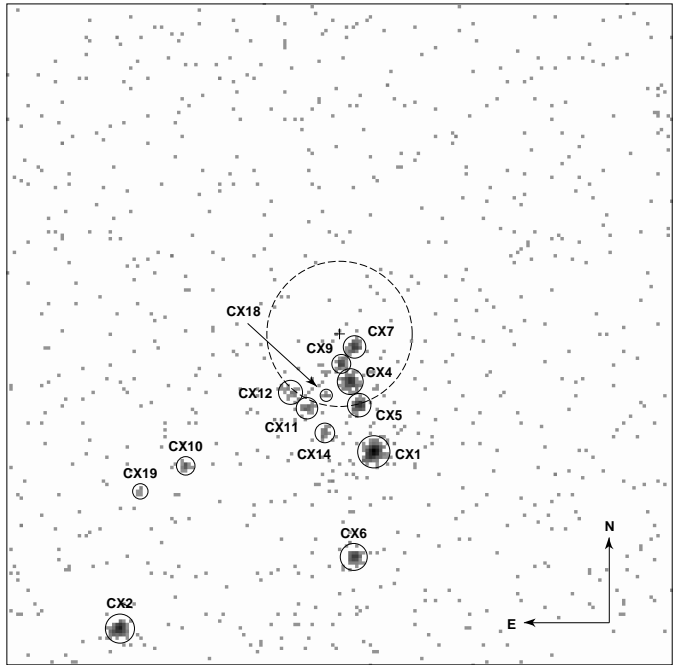}
\caption{X-ray image of the central $1\farcm6 \times 1\farcm6$ region
of NGC~6752.  The solid circles indicate the source extraction regions
as determined by {\tool wavdetect}.  The optical cluster center is
indicated with a cross, and the dashed circle is the 10\farcs5 core
radius of the cluster.  Each pixel is a 0\farcs492 square.\label{fig:xraycenter}}
\end{figure*}

\subsection{X-ray Source Detection} \label{sect:xraysrcs}
The wavelet-based {\tool wavdetect} tool was employed for source
detection.  We found that the detection of sources was insensitive to
whether we used a medium-band (0.5--4.5~keV) or full-band
(0.3--8.0~keV) image.  We detected 19 point sources within the cluster
half-mass radius (Table~\ref{tab:xraysrcs}) and another 21 on the rest
of the S3 chip.  We have numbered the sources in order of most counts
to least in the 0.5--6.0~keV band.  Our detection threshold was
$\geq$5 counts in the 0.3--8.0~keV image.  The absorbed flux of the
faintest detected source was $\sim6\times 10^{-16}$~\ergcms.  Using
the density of sources outside the half-mass radius, we estimate that
about four sources within the half-mass radius are not associated with
the cluster, which is in approximate agreement with the $\log{N}-\log{S}$
relationships of \citet{gia01}.  All 19 possible cluster sources are
consistent with being point sources, with the possible exception of
CX12.  Comparing the observed radial surface brightness profile of this
source to one predicted for a source of its intensity
and at its location on the chip, we find that it is inconsistent with
being a single point source at 94\% confidence ($\chi^2=12.2$ for 6
degrees of freedom).  It is most likely a blend of multiple sources.

The central 1\farcm6 portion of NGC~6752 is shown in
Fig.~\ref{fig:xraycenter}.  Overlaid on this image are the source
extraction regions determined by {\tool wavdetect}, a dashed circle
indicating the core radius, and a cross indicating the optical center
of the cluster.

\subsection{X-ray Count Rates} \label{sec:cts}
Source counts were extracted in a variety of energy bands.  We found
that the most useful data were between 0.5 and 6.0~keV, as this range
preserved almost all of the source counts while limiting the
background contribution.  We focus on the following bands for hardness
ratios and the X-ray color-magnitude diagram (CMD): 0.5--1.5~keV (\xsoft),
0.5--4.5~keV (\xmed), and 1.5--6.0~keV (\xhard).  While somewhat
arbitrary, this choice allows a direct comparison with the results of
\citet{gr01a} on 47~Tuc, which has a similar distance and column density.

The detected count rate was corrected for background, exposure
variations, and foreground photoelectric absorption.  The background count rate in each
band was estimated from an annulus around the innermost sources. The
inner radius was 29\arcsec, and the outer radius was 70\arcsec.  No
sources were present within the annulus.  Because of the spacecraft's
dither, some sources passed in and out of bad columns on the CCD.  To
account for the $\sim$4\% variations in exposure that resulted, we
applied multiplicative corrections based on the ratio of a source's
average effective area in each of the three bands to the average
effective area in the same band of CX15, which had the highest average
exposure.  The individual effective area curves for the sources were
made using the CIAO tool {\tool mkarf}.  The average effective area of
CX15 in each of the bands was 591~\cmsq\ (\xsoft), 465~\cmsq\ (\xmed),
and 377~\cmsq\ (\xhard).

\begin{figure*}[t!]
\plotone{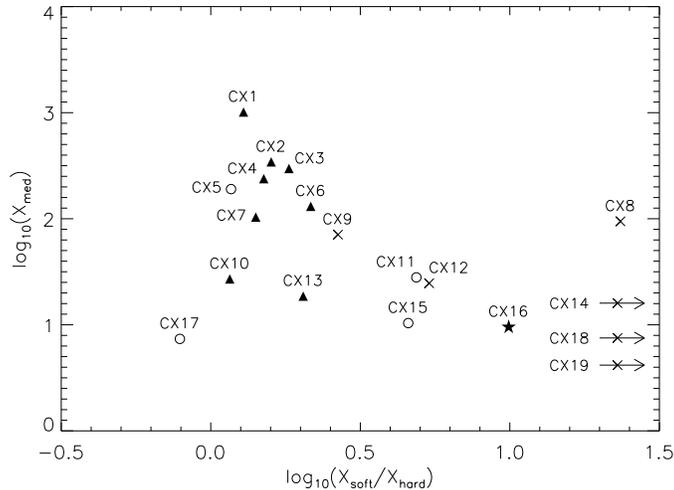}
\caption{X-ray color-magnitude diagram for NGC~6752. The X-ray color is defined as
the logarithm of the ratio of \xsoft\ (0.5--1.5~keV) corrected counts
to \xhard\ (1.5--6.0~keV) corrected counts, and the magnitude is the
logarithm of \xmed\ (0.5--4.5~keV) corrected counts.  Probable source
natures are indicated by symbols ($\blacktriangle$ for CVs,
$\bigstar$ for BY~Dra, {\LARGE $\circ$} for ambiguous sources with optical
or radio counterparts, and $\times$ for sources with no
radio or optical counterparts).  The arrows
indicate the magnitudes of the three sources (CX14, CX18, and CX19)
for which there were no detected counts in \xhard.
\label{fig:xraycmd}}
\end{figure*}

To correct for the column density to NGC~6752, we used \xspec\
\citep{arn96} to examine the effects of absorption on three spectra
characteristic of what one might expect to find in a globular cluster:
a 3~keV thermal bremsstrahlung (for CVs; Richman 1996; van Teeseling,
Beuermann, \& Verbunt 1996), a 0.3~keV blackbody (for quiescent LMXBs; Verbunt
et al.\ 1994; Asai et al.\ 1996, 1998; as substitute for the more
correct neutron-star atmosphere models, see Rutledge et al.\ 2001b and
references therein), and a power law with a photon
index of $\Gamma=2$ (for MSPs; Becker \& Tr\"{u}mper 1999).  In each
of our three bands, we took the ratio of the unabsorbed count rate to
one absorbed by a column density of $\nh = 2.2\times10^{20}$~\pcmsq.
Because this is a fairly low column density, the effects were small on
each of the three model spectra.  We used a simple
average of the ratios to derive the following correction factors for
the observed count rate in each band: 1.084 (\xsoft), 1.067 (\xmed),
and 1.010 (\xhard).  Table~\ref{tab:xraysrcs} lists both the observed
and fully corrected counts in each band for all 19 sources.  The X-ray
color-magnitude diagram based on the corrected counts is shown in
Fig.~\ref{fig:xraycmd}.  Note that the effect on the X-ray
color-magnitude diagram of the correction for absorption is a uniform
shift of all points 0.03 units up and 0.02 units to the right.

\subsection{X-ray Spectral Fitting}
We used the CIAO tool {\tool dmextract} to extract spectra of the
brighter sources (CX1--CX9) and {\tool dmgroup} to group the spectra
to a selected number of counts per bin.  CX1 was grouped to contain
$\geq$35 counts per bin, CX2--CX4 were grouped at $\geq$15 counts per
bin, and CX5--CX9 at $\geq$10 counts per bin.  Background-subtracted
spectral modeling was performed with \xspec\ using data in the
0.3--8.0~keV range.

We fit three different models to each of the nine brightest sources
(CX1--CX9): thermal bremsstrahlung (TB), blackbody (BB), and power law
(PL).  For CX1, the column density was allowed to vary and compared to
a fixed value of $\nh=2.2\times10^{20}$~\pcmsq\ for the cluster.  For
the other sources, however, we found that the data could not constrain
\nh\ well and that all best-fit values were consistent with the column
density from optical reddening.  We therefore fixed \nh\ at the
optical reddening value for the rest of the sources.  The results are
listed in Tables~\ref{tab:xrayspec} and~\ref{tab:cx1spec}.

From the best-fit models, we calculated the unabsorbed source
luminosities for CX1--CX9.  Averaging the results from the best fits
of each of the three models, we arrived at our estimate for the source
luminosities listed in Table~\ref{tab:xraysrcs}.  The major sources of
uncertainty in these luminosities arise from the scatter in the
luminosities from the three different models and the uncertainties in
each individual fit. The combined effect is typically around 20\%. The
uncertainties in distance (2.4\%) and column density (1.7\%) are
negligible in comparison.

Fitting a linear relation to the CX1--CX9 luminosities vs.\ the
corrected counts in \xmed, we have estimated the unabsorbed
luminosities for sources CX10--CX19 based on their \xmed\ counts.
These are also listed in Table~\ref{tab:xraysrcs}.

\section{Radio Observations} \label{sect:radio}
We have analyzed archival radio observations of NGC~6752, carried out
on 1995 February 02 using the Australia Telescope Compact Array (ATCA;
Frater, Brooks, \& Whiteoak 1992).  Observations were made
simultaneously at 1.4 and 2.4~GHz, and cover the entire \chandra\
field-of-view. At 1.4 and 2.4~GHz, the resulting spatial resolutions
are $5\farcs7 \times 4\farcs9$ and $3\farcs2 \times 2\farcs7$ with
sensitivities of 70~\uJy~beam$^{-1}$ and 90~\uJy~beam$^{-1}$,
respectively (where 1~Jy~$=10^{-26}$~W~m$^{-2}$~Hz$^{-1}$).

We find three source coincidences between the X-ray and 1.4-GHz radio
images: one point source within NGC~6752's half-mass radius (CX17), one
point source $\sim6'$ from the cluster center (the corresponding X-ray
source falls on chip S4), and some diffuse X-ray emission centered on a
bright extended radio source (possibly a head-tail radio galaxy)
outside the cluster. At 1.4~GHz the radio counterpart to CX17 has a
flux density of $0.5\pm0.1$~mJy; no circular or linear polarization is
detected from this source, with a $3\sigma$ upper limit on the
fractional polarization of 35\%.  At 2.4~GHz this source is marginally
detected at the $3\sigma$ level, with a flux density $\sim0.3$~mJy.

The absolute astrometry of the ATCA data in the International
Celestial Reference System (ICRS) is accurate to $\la0\farcs1$.
Based on the two point-source coincidences, we corrected the
\chandra\ astrometry to align these sources. The correction
was $-0\fs044$ ($=-0\farcs33$) in RA and $-0\farcs17$ in Dec.

\section{Optical Observations} \label{sect:optid}
The {\it HST} archive contains a substantial amount (\til 400
exposures) of WFPC2 imaging data covering the central regions of NGC
6752.  Of particular utility are the data obtained by Bailyn and
collaborators to suggest optical identification for two \rosat\
sources in the core, the very ones we have now resolved into multiple
\chandra\ sources.  These consist of deep $H\alpha_{656}$ (F656N
filter; i.e., 6560\AA), $R_{675}$ (F675W) and $B_{439}$ (F439W) image
sets, plus long time-series in $V_{555}$ (F555W) and $I_{814}$ (F814W)
covering the core alone (for details see Bailyn et al.\ 1996), ideal
for the identification of H\,$\alpha$ emission and/or variable stars.
Furthermore, additional $U_{336}$ (F336W), $nUV_{255}$ (F255W) are now
available allowing for additional UV-excess selection.  Unfortunately,
not all of our {\it Chandra} sources have such complete optical
coverage, owing to the nature of this archival data obtained for other
unrelated projects.  However, only one X-ray source (CX8) lacks any
\hst\ color information at all.

In this section, we will first outline the optical astrometry
undertaken to tie the {\it HST} images to the ICRS and thereby
facilitate the alignment of our \chandra\ detections.  We will then
proceed to describe our optical data preparation and reductions, and
end with our results.

\subsection{Optical Astrometry} \label{sect:optastrom}
We aim to tie each {\it HST} pointing to the ICRS by finding matches
between stars appearing on {\it HST} images and stars with accurate
positions in either the Tycho-2 \citep{hog00} or USNO A-2
\citep{mon98} catalogs.  Unfortunately, owing to the brightness and
stellar density of the cluster core neither catalog extends to within
5\arcmin\ of the center.  However, the {\it HST} archive also contains
numerous (largely V-band) images extending from the cluster center out
to a \til10\arcmin\ radius, and these outermost images do overlap with
the USNO A-2 coverage.  On the basis of the \hst\ pointing information
(contained in each image header), we used the IRAF/STSDAS routine
{\tool metric} (which includes corrections for geometrical distortions
towards the edge of each chip) to over-plot the nominal USNO star
positions.  To aid in the matching we also then degraded the \hst\
images by convolving with a Gaussian to approximate the typical
ground-based seeing of the plates used in deriving the catalog.  It
was then possible to make 58 reliable matches, on two separate \hst\
datasets (i.e., from different satellite pointings), deriving offsets
of 1\farcs1 and 0\farcs1 in RA and Dec., respectively. We found no
evidence for a significant discrepancy in the nominal roll-angle, and
hence we made no correction to this parameter.  Taking the scatter to
the astrometric fit and, in particular, the uncertainty in the
roll-angle, we estimate residual uncertainties of a few tenths of an
arcsecond.

Having aligned these outermost \hst\ images, we then had to
step in via intermediate overlapping \hst\ images to find corrections
for all the useful images of the cluster center.  The procedure was
essentially the same for each step.  We measured the centroids of all
bright (but not saturated) stars appearing in the overlap regions of
two \hst\ datasets and then applied {\tool invmetric} to derive their
nominal RA and Dec according to the headers.  Pairs were then matched
and average offsets in sky coordinates calculated.  Again, in no case
did there appear to be any significant roll-angle discrepancies
present.  For these steps, we more typically found a few hundred
matches, and hence the uncertainty in the measured offsets was
negligible relative to that of the initial tie to the USNO A-2 (ICRS)
frame.

The final stage made use of the prior identifications of two
H\,$\alpha$ bright stars by \citet{ba96}.  Over-plotting our \chandra\
sources for the core revealed that the two optical stars flagged in
the Bailyn et al.\ finding chart (their Fig.~2) are indeed excellent
matches to two of our X-ray sources. Specifically, our analyses of the
relative positions and position angles in both the X-ray and the
optical for these two flagged stars were fully in agreement. (Note
that our astrometric solution for the position of these two objects
differs in RA and Dec by 0\fs33 ($=2\farcs6$) and 8\farcs4,
respectively, from the positions given by Bailyn et al.\ 1996.  We can
suggest no explanation for this large discrepancy.) In our analysis,
only a small \til0\farcs5 X-ray vs.\ optical offset remained.
However, this is consistent with the expected uncertainty in our
astrometric solution for these \hst\ images relative to the ICRS
frame.  Hence, we decided to apply this small offset (assuming that
these two stars are exactly at their \chandra\ positions) to the
astrometry for all of our \hst\ datasets. In order to be thorough, we
opted to examine all optical/UV stars within \til0\farcs5 of each
\chandra\ source position. Indeed, as we describe in detail below, we
then uncovered at least 3 more strong candidate optical counterparts,
none of which was more than 0\farcs1 from the expected position based
upon this offset.  Such results are indeed consistent with
uncertainties expected in the relative astrometry of the X-ray
positions, combined with their transfer to a variety of \hst\ datasets.

\subsection{Optical Data Reductions and Analysis}
The data from the \hst\ archive are already well-calibrated (aside
from astrometry), and only two steps remained.  For each pointing and
filter, at least two images were available, which were then used to
remove the numerous cosmic ray events detected in each exposure.  We
used the {\tool combine} task to calculate an average image and
applied one-sided sigma clipping (based on the median value of each
pixel and a noise model for the detector) to exclude cosmic ray
events. The images from each of the four chips were then trimmed to
remove the partially masked edge regions.

We employed {\tool daofind} in the {\sc IRAF} implementation of {\sc
DAOPHOT II} \citep{stet92} to find all stars detected at approximately
3$\sigma$ above background, and we then determined their centroids.
Unfortunately, this method is not fool-proof, and we also had to add a
number of faint stars by hand to our list and delete others affected
by the \til20 saturated stars appearing on each frame.  We excluded
all stars with centers within \til10--15 pixels of any saturated
pixels and also any stars lying on the cross-shaped diffraction
patterns of the very brightest stars.  Having thus created a master
star list for each image, aperture photometry yielded magnitudes for a
range of annuli close to the measured FWHM of the PSF.  We note that,
for the undersampled stellar PSFs of the PC and especially the WF
chips, aperture photometry produced better results than our attempts
at PSF fitting.  The apertures we used are, however, far smaller than
the standard 0\farcs5 aperture of the STMAG system \citep{holt95}, and
we therefore estimated the corresponding magnitude offsets needed
relative to this large aperture for a number of bright (but
non-saturated) and isolated stars on both the PC and WF chips.
Applying both these offsets and the appropriate zeropoints based upon
the sensitivity information in each header, we finally arrived at the
STMAGs in a given filter for each star.  We note that accurate
estimation of the background contribution (especially critical for the
faintest stars), is very difficult in such a crowded field.  We
adopted the mode value for an annulus surrounding each star.  Tests,
using either a fixed value as determined by averaging the results for
the least crowded stars (farther from the core) or using the centroid
of a Gaussian fit to the ``sky'' histogram instead of the mode, showed
that we can expect systematic errors of \til0.1--0.2 mags in our final
STMAG values.  We have also applied approximate corrections to the
magnitudes derived from aperture photometry to compensate for the
time-dependent charge-transfer-efficiency effect and the so-called
``long versus short anomaly,'' according to the formulae presented on
the WFPC2 instrument web-pages
(\url{http://www.stsci.edu/instruments/wfpc2/}).

\begin{figure*}[!htb]
\resizebox*{!}{0.8\textheight}{\rotatebox{0}{\plotone{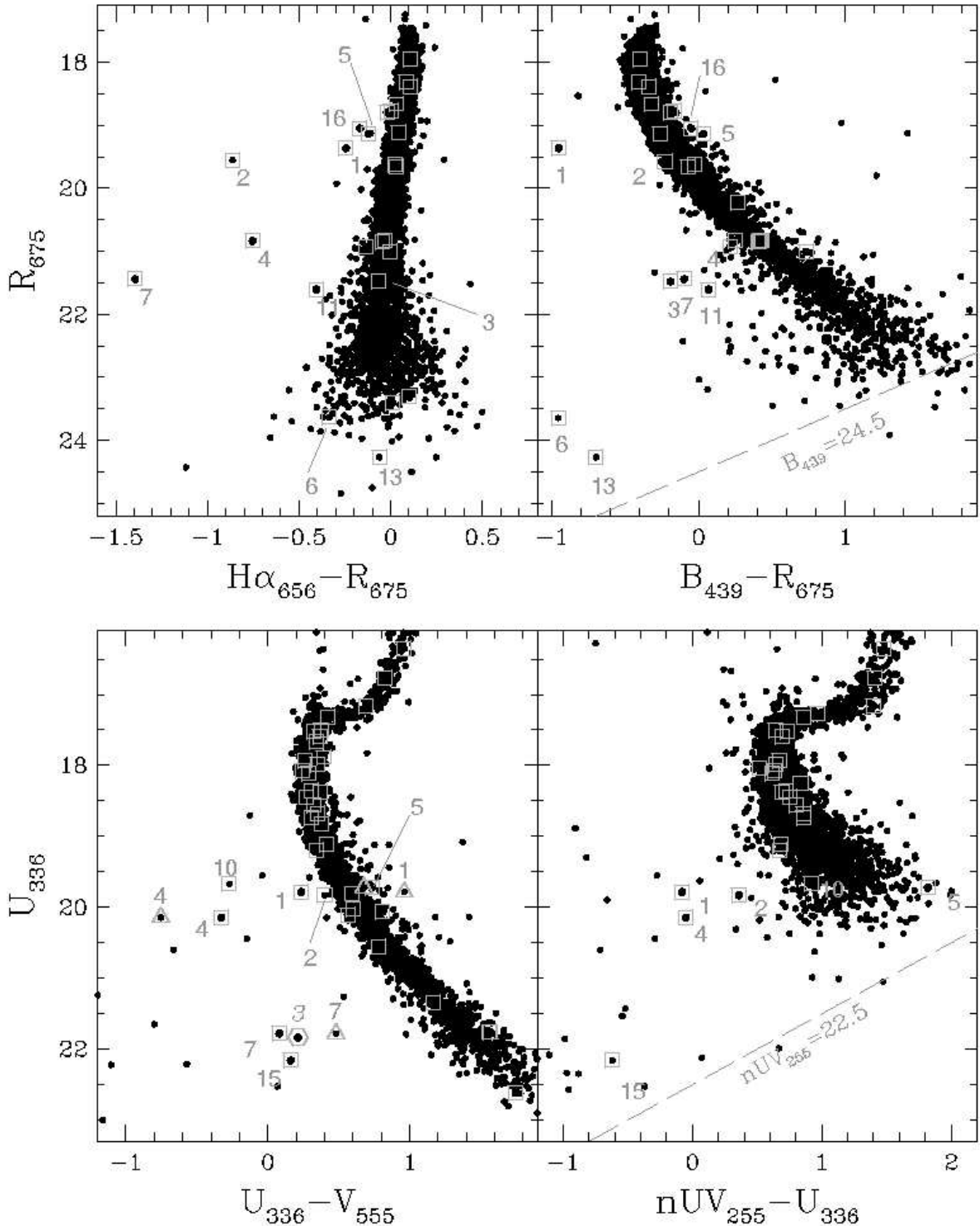}}}
\caption{Color-magnitude diagrams for the central regions of NGC~6752.
All the stars from both the PC field and one of the WF chip fields are
plotted, along with all stars within 0\farcs5 of the \chandra\ source
positions (indicated by grey boxes).  Numbers refer to the candidate
optical counterparts to the X-ray sources.  All the $R_{675}$,
$H\alpha_{656}$, and $B_{439}$ band data shown in the upper panels are
taken from \hst\ observations in 1994 August, whereas the $U_{336}$,
$V_{555}$ and $nUV_{255}$ data in the lower panels were obtained in
2001 March.  Note that in the bottom left panel: (i) four candidates
are plotted twice, according to their colors calculated from $V_{555}$
band data taken either contemporaneously (i.e., that of 2001 March,
squares) or older data from 1994 August (triangles), which clearly
shows that three of them are variable; (ii) the data for Star 3
(marked as a hexagon) were obtained in the wide $U$ (F300W) and wide
$V$ (F606W) filters and have been transformed before
plotting. \label{fig:CMDs}}
\end{figure*}

\begin{figure*}[!htb]
\resizebox*{!}{0.8\textheight}{\rotatebox{0}{\plotone{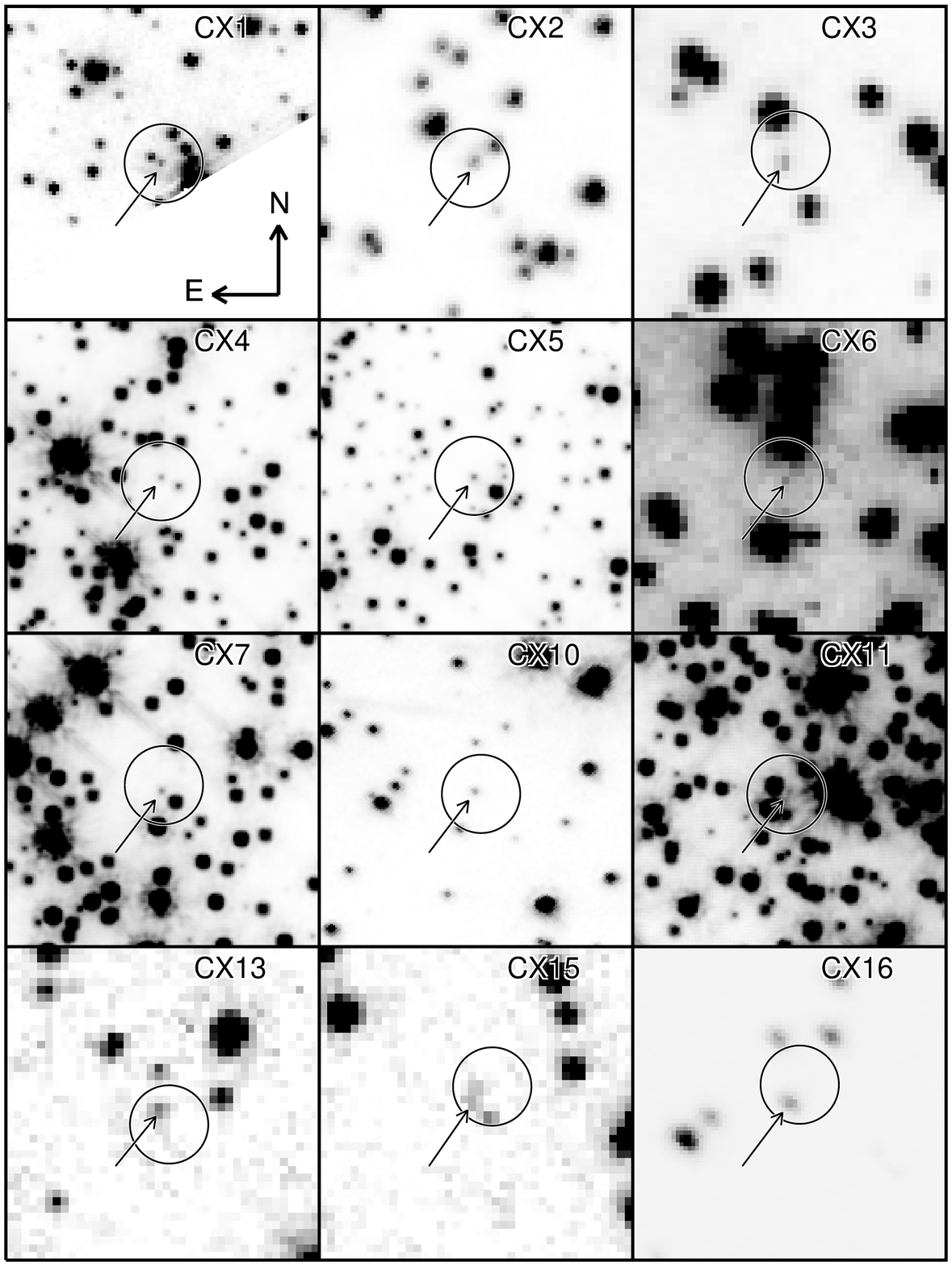}}}
\caption{4\arcsec$\times$4\arcsec\ finding charts for each optical
counterpart candidate, obtained from \hst\ archival data. These images
were taken in $V_{555}$, apart from the charts for Stars 1
($U_{336}$), 3, 6, and 13 ($B_{439}$), and 14 ($U_{336}$), for which
$V_{555}$ was either unavailable or the counterpart too blue or
crowded to be visible.  We have overlaid the 0\farcs5 radius error
circles for the \chandra\ source positions (in which we searched), and
the candidate stars themselves are indicated by arrows.  We also note
that both the depth, greyscaling and pixel scales for the images are
varied; as we have chosen the deepest, most well-sampled images (using
dithered images where available) and then adjusted the greyscale to
enhance visibility of the candidates.  \label{fig:fcharts}}
\end{figure*}

We constructed a variety of color-magnitude diagrams (CMDs) with all
the available data.  The most informative of these diagrams are shown
in Fig.~\ref{fig:CMDs}, on which {\it all} stars located within the
0\farcs5 \chandra\ error circles are indicated by grey
boxes. Numbers have been assigned to all candidate counterparts
corresponding to the ``CX'' designation (i.e., our Star 1 is the
counterpart to CX1).  In this scheme, Stars 1 and 2 of \citet{ba96}
are CX4 and CX7, respectively.  In some cases, candidates are
only abnormal in certain colors, but they are all labelled in each
diagram.  The results on each star are summarized in
Table~\ref{tab:res} and \S\ref{sect:res}, and finding
charts are shown in Fig.~\ref{fig:fcharts}.

\section{Results} \label{sect:res}
The X-ray luminosities for the \chandra\ sources as listed in
Tables~\ref{tab:xraysrcs} and \ref{tab:xrayspec} are in a range
covered in the galactic disk by CVs, RS CVn binaries, and MSPs (e.g.,
see Fig.~8 in Verbunt et al.\ 1997).  The soft X-ray transients with
neutron stars in the galactic disk tend to have higher luminosities
(the lowest luminosity known to us is that of Cen~X-4 measured in
August 1995 with the \rosat\ HRI at \ee{7}{31}~\ergsec\ between 0.5
and 2.5~keV; Rutledge et al.\ 2001).  Transients with a black hole
have quiescent luminosities down to $\sim$10$^{30}$~\ergsec; these
luminosities may be due to chromospheric emission from the companion
to the black hole, however \citep{rut00}.  None of the X-ray spectra
in our sample is as soft as would be expected for a low-mass X-ray
transient in quiescence; all systems with enough signal (CX1--9),
except CX8, can be fit with relatively hard thermal bremsstrahlung
spectra, as expected for CVs.

\begin{figure*}[t]
\resizebox*{\textwidth}{!}{\rotatebox{-90}{\plotone{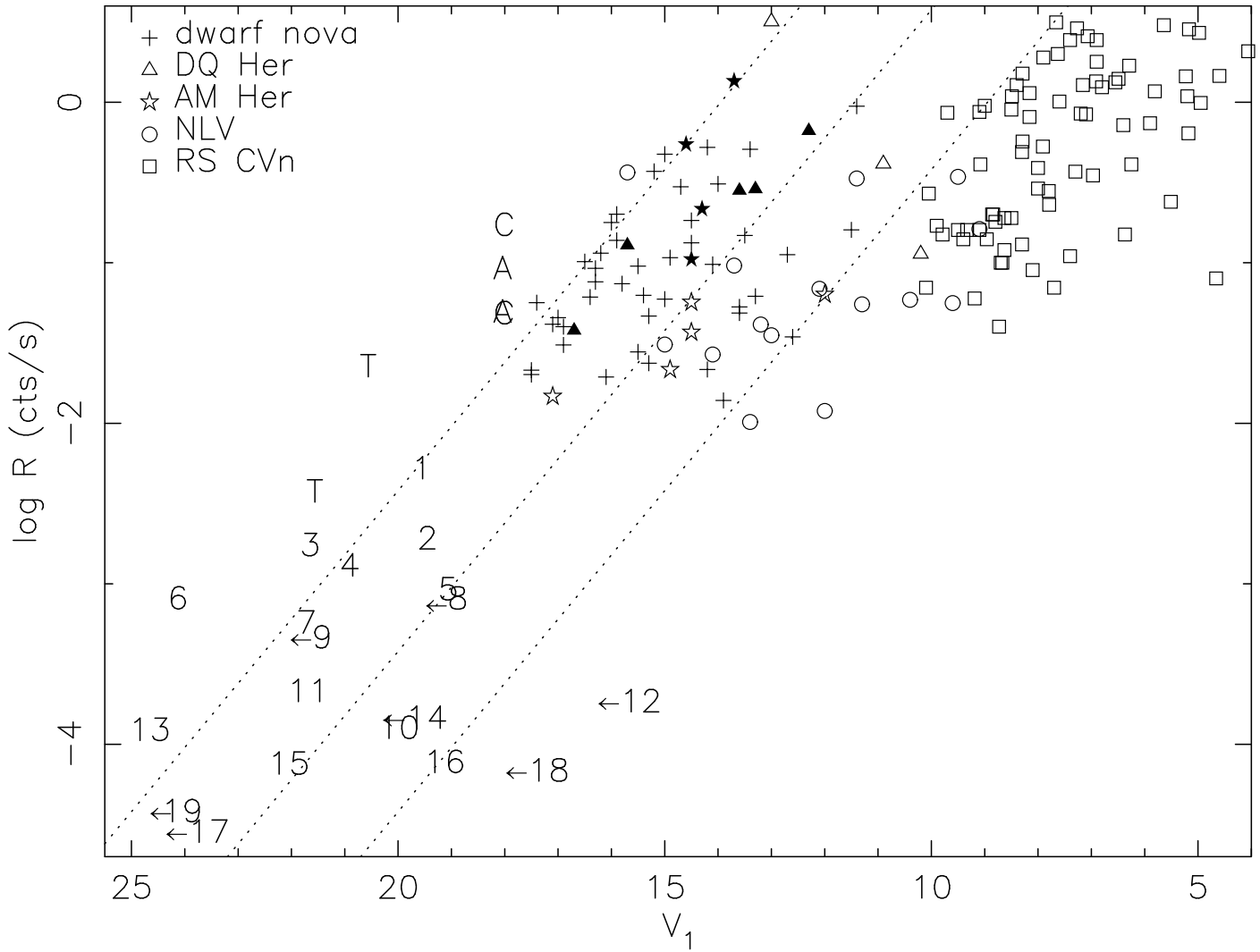}}}
\caption{Comparison of the X-ray and optical fluxes of \chandra\
sources (numbers) in NGC~6752 to those of field sources (symbols and
letters --- ``C'' for Cen X-4, ``A'' for Aql X-1, and ``T'' for X9/V1
and X19/V2 in 47~Tuc; see Verbunt \& Hasinger 1998 for more details).
$R$ is the equivalent \rosat\ countrate in channels 52--201.  $V_1$ is
the visual magnitude and for CVs is the magnitude at which the system
is most commonly found.  \chandra\ sources with only optical upper
limits are indicated by arrows. The top dotted line roughly separates
soft X-ray transients (above) from CVs (below).  The bottom dotted
line roughly separates CVs from RS~CVn systems (below).
\label{fig:xrayopt}}
\end{figure*}

A further clue to the nature of the \chandra\ sources may be obtained
from the X-ray to optical flux ratio.  For the \chandra\ sources whose
optical counterparts have not been measured in $V$, we estimate
$V\simeq0.5(B+R)$.  To allow comparison with the large database of
\rosat, we estimate that one \xsoft\ \chandra\ count for our 29~ks NGC~6752
observation corresponds to a rate of \ee{0.9}{-5}~counts~s$^{-1}$ in
channels 52--201 of the \rosat\ PSPC.  As shown in
Fig.~\ref{fig:xrayopt}, the ratio of optical to X-ray flux of most
Chandra sources in NGC~6752 suggests that they are cataclysmic
variables. We discuss these first and the possible exceptions of CX12,
CX16, and CX18 later.

Fig.~\ref{fig:CMDs} shows that the optical counterparts to CX1, CX2,
CX3, CX4, CX7, CX10 and CX15 all have ultraviolet excesses (at
$U_{336}$ and/or $nUV_{255}$) with respect to the main sequence, as
expected for cataclysmic variables.  In $B_{439}-R_{675}$, the optical
counterparts to CX1, CX3 and CX7 are still blue with respect to the
main sequence, CX2 and CX4 are not distinguishable from the main
sequence, and no information is available on CX10 and CX15.  CX6, CX11
and CX13, for which no ultraviolet fluxes are available, are bluer
than the main sequence in $B_{439}$.  CX1, CX2, CX4, and CX7 also show
clear H\,$\alpha$ emission, and CX3, CX6 and CX13 do not.  CX11 shows
marginal H\,$\alpha$ excess, and no information is available for CX10
and CX15.  On the whole, the optical properties for all these objects
are compatible with those expected for cataclysmic variables, although
CX11 and CX15 show evidence for having spatially extended emission and could be
background galaxies. These and other early \chandra\ observations are
finding the long-sought but, until recently, rarely observed population of
CVs in globular clusters \citep{gr01a,gr01b}.

CX12, CX18, and CX16 have (limits to) X-ray to optical flux ratios in
the range of RS~CVn systems.  The limited information that may be
obtained from the X-ray hardness of these three sources indicates that
they may be as soft as expected for RS~CVns.  Interestingly, the $B_{439}-R_{675}$
color of CX16 places it on or slightly above the main sequence (see
Fig.~\ref{fig:CMDs}); its weak H\,$\alpha$ emission is not
unprecedented in RS~CVn type systems either.  The optical counterparts
for CX12 and CX18 would have visual fluxes more than about 3 mag
fainter than our limits if they were dwarf novae in quiescence. They
can be as bright as our limits if they are dwarf novae in outburst or
nova-like variables of the UX UMa type (i.e., non-magnetic,
permanently bright cataclysmic variables).

The location of CX5 in the optical color-magnitude diagrams is
similar to that of CX16, as is its moderate H\,$\alpha$ emission. This
suggests that it is an RS~CVn system (more precisely, a BY~Dra
system). Its high X-ray to optical flux ratio and
spectral hardness are more suggestive of a cataclysmic variable,
however.

As remarked above, our \chandra\ sources have X-ray luminosities in
the range also observed for MSPs in the galactic disk.  In this
respect, the X-ray sources for which we have no optical counterparts
could be MSPs.  Also, as stated in \S\ref{sect:xraysrcs},
we expect $\sim$4 serendipitous background sources within the
half-mass radius. The optically unidentified faint X-ray sources may
well be such background sources.  Alas, no period derivatives or
positions are as yet available for the five MSPs, with periods ranging
from 3.3 to 9.0~msec, now known in NGC~6752 \citep{pos01}. The X-ray
luminosity of an MSP is only a small fraction of the spin-down luminosity
\citep{vb96,pos01b}, and the latter is proportional to $P^{-3}$ (with $P$ the
pulse period).  To illustrate this we write the X-ray luminosity
$L_{\rm x}$ as a fraction $f$ of the spindown luminosity $\dot E_{\rm
sd}$
\begin{equation}
L_{\rm x}=f\dot E_{\rm sd}
         ={4\pi^2If\dot P\over P^3}
         ={2\pi^2If\over\tau_{\rm c}P^2}
\end{equation}
where $I\simeq 10^{45}$g~cm$^{2}$ is the moment of inertia of the 
neutron star,
$\tau_{\rm c}\equiv P/(2\dot P)$ its characteristic age, and $\dot{P}$
the intrinsic period derivative. 
Entering characteristic values we obtain
\begin{equation}
L_{\rm x}= 7\times 10^{30}\,{\rm erg s}^{-1} {f\over 0.001}
{10^{10}{\rm yr}\over\tau_{\rm c}}\left({0.003}\,{\rm s}\over P\right)^2.
\end{equation}
We think that the most likely candidates are PSRs J1910$-$59A, C and
E, which have the shortest periods and should therefore be the
brightest in X-rays. The least luminous X-ray sources appear the more
likely candidates, in particular CX17, which has a radio counterpart.
However, CX8 has a very soft spectrum, and in our view cannot be
excluded.

All of our likely source identifications are indicated in the X-ray
color-magnitude diagram (Fig.~\ref{fig:xraycmd}).

\section{Discussion} \label{sect:dis}
Our \chandra\ observation provides a dramatic improvement over the
\rosat\ HRI observations. The central X-ray emission of NGC~6752,
resolved into 4 sources by the \rosat\ HRI, is now found to be due to
ten sources.  The very accurate positions of these and other newly
discovered sources allow secure identifications with \hst\ sources
and with one radio source.

Outside the core, the two sources CX2 and CX3 correspond to the
\rosat\ sources X14 and X6, respectively. From these two we determine
an offset of (1\farcs9, $-$2\farcs1) that has to be applied to the
\rosat\ positions (as given by Verbunt \&\ Johnston 2000) to align
them with the \chandra\ frame.
 
\begin{figure*}[!htb]
\plotone{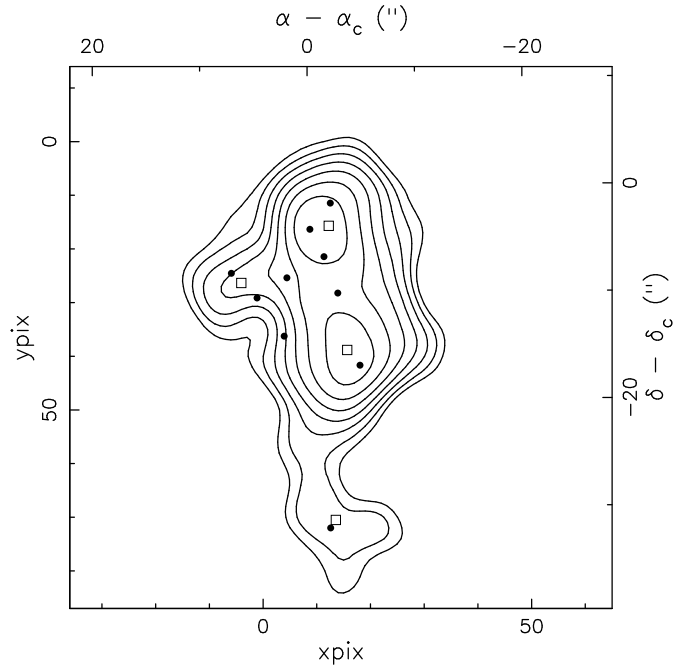}
\caption{\rosat\ HRI contours of NGC~6752 with positions of the central \chandra\
sources superposed (solid circles). \rosat\ centroids are also
indicated (boxes).  The \rosat\ contours were shifted (based on the
correspondance of CX2 and CX3 with X14 and X6, respectively, from
Verbunt \& Johnston 2000) to align them with the \chandra\ frame.  \label{fig:rosatcontours}}
\end{figure*}

In Fig.~\ref{fig:rosatcontours} we show the positions of the central
sources that we have detected with \chandra\ superposed on a contour
of the \rosat\ HRI observation to which this offset has been applied.
We then find that CX6 and CX1 correspond to X22 and X7b, respectively,
whereas \rosat\ source X7a has been resolved into CX11, CX12, and
CX14, and X21 into CX4, CX5, CX7, and CX9.

From our spectral fits to the \chandra\ sources, we can estimate the
\rosat\ HRI and PSPC countrates for the luminosities at which
\chandra\ detected them: 1 \xsoft\ \chandra\ count in our observation
corresponds to about 0.0045 cts/ks in the \rosat\ HRI observations
and to 0.009 cts/ks (channels 52--201) in the \rosat\ PSPC
observation.  With this conversion we find that the luminosities of
X14/CX2, X6/CX3, X22/CX6, X7a/(CX4, 5, 7, 9) and X21/(CX11, 12, 14)
are constant within the errors between the \rosat\ HRI and \chandra\
observation.  Source X7b/CX1, however, was about 30\%\ brighter in the
\rosat\ HRI than in the \chandra\ observation.  The central
conglomerate of sources was about 40\%\ brighter during the \rosat\
PSPC observation than during the \chandra\ observation; due to the
lack of spatial resolution in the PSPC we cannot assign this to a
specific \chandra\ source.  We thus have evidence that the X-ray
luminosity of at least one central source, CX1, is variable.  Both
X-ray and optical variability will be discussed in detail in
\citet{hom01b}.
 
\chandra\ observations have been published for three other globular
clusters that do not contain a bright ($L_{\rm x}>10^{36}$~\ergsec)
central X-ray source. The numbers of {\em identified} sources of
various types are summarized in Table~\ref{tab:compare}.  Of these
clusters, 47~Tuc has a larger core and slightly smaller central
density than NGC~6752, causing the frequency of close encounters in
47~Tuc to be a factor $\sim$10 larger than that in NGC~6752.  NGC~6397
has a higher core density and much smaller core radius than NGC~6752,
leading to a frequency of close encounters a factor $\sim$10 smaller
than in NGC~6752.  Finally, $\omega$~Cen has a smaller central density
and larger core than any of these clusters, and an encounter frequency
similar to that in NGC~6752.

The X-ray luminosity of currently known quiescent soft X-ray
transients with neutron stars in the Galactic disk does not drop below
$\sim10^{32}$ erg~s$^{-1}$. If a similar lower limit holds for such
transients in globular clusters, the \chandra\ observations provide a
complete census of the quiescent transients in the observed four
clusters, since none of the unclassified sources has an X-ray
luminosity in excess of this threshold.  If the number of quiescent
soft X-ray transients scales with the collision frequency, we estimate
that one such system is formed per $\sim$30 of our normalized
encounter frequencies, in which case the numbers of two, one, and zero
in 47~Tuc, $\omega$~Cen, and NGC~6752 are well within expectation with
Poisson statistics.  The possible presence of such a source in
NGC~6397, however, is somewhat surprising.  A similar statement may be
made for the X-ray detected MSPs: for an expected number of one per
ten normalized encounter frequencies, only NGC~6397 is surprising in
containing one.  The presence of both a soft X-ray transient and an
X-ray detected MSP in NGC~6397 suggests that clusters with the highest
densities contain more such systems than indicated by the (average)
collision number.  This may be the consequence of more pronounced mass
segregation in these clusters, which enhances the encounter rate of
neutron stars with respect to the average encounter rate by
concentrating them to the core.

The relative numbers of cataclysmic variables and binaries with a
neutron star (i.e., permanently bright low-mass X-ray binaries or soft
X-ray transients) according to theory depends on a number of factors
including the retention fraction of neutron stars (many neutron stars
may be born with velocities that throw them out of the cluster,
whereas all white dwarfs remain), the mass segregation in the cluster
(which will concentrate the remaining neutron stars to the core where
the close encounters occur, more so than the less massive white
dwarfs), and the importance of primordial binaries evolving into
cataclysmic binaries (important in clusters with small encounter
frequencies).  It is therefore interesting to see that the number
ratio of (hitherto classified) cataclysmic variables to soft X-ray
transients is not significantly different in 47~Tuc and NGC~6397, even
though their encounter frequencies differ by two orders of magnitude.
If many of the unclassified sources in 47~Tuc turn out to be
cataclysmic variables, as is expected, the number ratio of cataclysmic
variables to soft X-ray transients in it will even be higher.  This
again suggests that the average collision number of NGC~6397
underestimates the neutron star encounter rate.  It should be noted
that the census of cataclysmic variables through X-rays is certainly
not complete, as many cataclysmic variables observed in the Galactic
disk have X-ray luminosities below our threshold of
$2\times10^{30}$~\ergsec\ (see e.g.\ Fig.~8 in Verbunt et al.\ 1997).
If we accept that X9/V1 and X19/V2 in 47~Tuc and CX6 in NGC~6752 are
indeed cataclysmic variables, we see from Fig.~\ref{fig:xrayopt} that
their X-ray to optical flux ratio is higher than that of any
cataclysmic variable in the galactic disk studied by \citet{vb96}. The
high X-ray luminosity of X9 has been confirmed with Chandra
observations; the relative hardness of its X-ray spectrum suggests a
cataclysmic variable, rather than a quiescent soft X-ray transient
\citep{gr01a}

Equally remarkable is the fact that the number ratios of cataclysmic
variables to X-ray active binaries is $\sim4$ in all four clusters
under discussion, notwithstanding the large range of encounter
frequencies. In dense clusters like 47~Tuc, most cataclysmic variables
should be formed via close encounters, whereas a less concentrated
cluster like $\omega$~Cen may have a significant contribution of
primordial binaries evolved into cataclysmic variables
\citep{vb88,di94,da97}.  As more sources are identified in 47~Tuc and
$\omega$~Cen, and other clusters are studied as well, the numbers may
become big enough to allow more definite conclusions.

Finally, we are struck by the fact that all X-ray sources in
Fig.~\ref{fig:xraycenter} lie in a quadrant south of the cluster
center. The formal probability of this happening with thirteen sources
is small, but, if the optical cluster center position is accurate, we
are inclined to ascribe this to chance.  At the small interstellar
absorption towards NGC~6752, and with the relatively hard spectra of
the X-ray sources, differential absorption over the cluster cannot
explain the asymmetry. The optical cluster center would have to be in
error by about 10\arcsec\ for it to be close to the X-ray center.  We
note that, if this is the case, then NGC~6752 becomes the second
globular cluster for which the optical determination of its center has
been corrected by an X-ray observation (the first one being NGC~6541,
see Fig.~1 of Verbunt 2001).

\acknowledgments
DP acknowledges that this material is based upon work supported under
a National Science Foundation Graduate Fellowship.
WHGL gratefully acknowledges support from NASA.
LH and SFA acknowledge the support of NASA through LTSA grant NAG5-7932.
The Australia Telescope is funded by the Commonwealth of Australia for
operation as a National Facility managed by CSIRO.  BMG acknowledges
the support of NASA through Hubble Fellowship grant HST-HF-01107.01-A
awarded by the Space Telescope Science Institute, which is operated by
the Association of Universities for Research in Astronomy, Inc., for
NASA under contract NAS 5--26555.
VMK is a Canada Research Chair and acknowledges support from LTSA
grant NAG5-8063, NSERC Rgpin 228738-00, and a Sloan Fellowship.

\begin{sidewaystable}
\caption{NGC~6752 X-ray Sources. \label{tab:xraysrcs}}
\begin{tabular}{lllllllll}
\tableline \tableline
\multicolumn{1}{c}{Src\tablenotemark{a}}& \multicolumn{1}{c}{RA (J2000)\tablenotemark{b}}& \multicolumn{1}{c}{Dec (J2000)\tablenotemark{b}}& \multicolumn{3}{c}{Detected
Counts/Corrected Counts\tablenotemark{c}} & \multicolumn{1}{c}{\Lx\
(\ergsec)\tablenotemark{d}} & \multicolumn{1}{c}{Cntrpt\tablenotemark{e}} & \multicolumn{1}{c}{ID\tablenotemark{f}}\\
\multicolumn{1}{c}{} & \multicolumn{1}{c}{} & \multicolumn{1}{c}{} &
\multicolumn{1}{c}{\xsoft} & \multicolumn{1}{c}{\xmed} & \multicolumn{1}{c}{\xhard} &
\multicolumn{1}{c}{[0.5--2.5~keV]} & \multicolumn{1}{c}{} &
\multicolumn{1}{c}{} \\
\tableline
CX1&  19:10:51.098& $-$59:59:11.83& 536/588.0& 936/1014.0& 444/457.0& \ee{2.1}{32}& Opt./X7b& CV\\
CX2&  19:10:55.972& $-$59:59:37.33& 196/215.3& 315/343.5 & 130/135.3& \ee{6.0}{31}& Opt./X14& CV\\
CX3&  19:10:40.318& $-$59:58:41.29& 177/195.4& 274/297.6 & 104/107.0& \ee{5.3}{31}& Opt./X6& CV\\
CX4&  19:10:51.546& $-$59:59:01.71& 132/146.0& 219/238.7 &  94/97.1 & \ee{4.0}{31}& Opt./B1& CV\\
CX5&  19:10:51.374& $-$59:59:05.11&  94/103.7& 175/190.3 &  86/88.7 & \ee{3.6}{31}& Opt.& CV/BY Dra\\
CX6&  19:10:51.462& $-$59:59:26.99&  83/90.9 & 121/130.9 &  41/42.1 & \ee{2.2}{31}& Opt./X22& CV\\
CX7&  19:10:51.467& $-$59:58:56.72&  58/64.3 &  95/103.6 &  44/45.4 & \ee{1.9}{31}& Opt./B2& CV\\
CX8&  19:11:02.944& $-$59:59:41.89&  83/90.8 &  87/94.5  &   4/3.9  & \ee{2.1}{31}& ---& \\
CX9&  19:10:51.720& $-$59:58:59.16&  47/52.0 &  65/70.8  &  19/19.5 & \ee{1.3}{31}& ---& \\
CX10& 19:10:54.706& $-$59:59:13.87&  13/14.2 &  25/27.0  &  12/12.3 & \ee{6.0}{30}& Opt.& CV\\
CX11& 19:10:52.376& $-$59:59:05.56&  22/24.1 &  26/28.0  &   5/5.0  & \ee{6.2}{30}& Opt.& CV/Gal.\\
CX12& 19:10:52.694& $-$59:59:03.27&  19/20.8 &  23/24.7  &   4/3.9  & \ee{5.6}{30}& ---& \\
CX13& 19:10:40.565& $-$60:00:06.07&  13/13.8 &  18/18.6  &   7/6.8  & \ee{4.6}{30}& Opt.& CV\\
CX14& 19:10:52.039& $-$59:59:09.12&  15/16.4 &  15/16.0  &   0/--   & \ee{4.2}{30}& ---& \\
CX15& 19:10:55.798& $-$59:57:45.53&   8/8.5  &  10/10.4  &   2/1.9  & \ee{3.2}{30}& Opt.& CV/Gal.\\
CX16& 19:10:42.474& $-$59:58:42.83&   8/8.7  &   9/9.5   &   1/0.9  & \ee{3.0}{30}& Opt.& BY Dra\\
CX17& 19:11:05.279& $-$59:59:04.03&   3/3.2  &   7/7.4   &   4/4.0  & \ee{2.7}{30}& Rad.& MSP/Gal.\\
CX18& 19:10:52.006& $-$59:59:03.69&   7/7.7  &   7/7.5   &   0/--   & \ee{2.7}{30}& ---& \\
CX19& 19:10:55.577& $-$59:59:17.54&   4/4.3  &   4/4.2   &   0/--   & \ee{2.2}{30}& ---& \\
\tableline
\end{tabular}

$^{\rm a}${Sources are numbered according to their total
counts.}

$^{\rm b}${The \chandra\ positions have been corrected by
$-0\fs044$ ($=-0\farcs33$) in RA and $-0\farcs17$ in Dec (see
\S\ref{sect:radio}).  We estimate the uncertainties in the final
astrometric solution at about 0\farcs3 in both RA and Dec.  These
uncertainties are much larger than the {\tool wavdetect} centroiding
uncertainties. }

$^{\rm c}${Corrections are described in \S\ref{sec:cts}.
X-ray bands are 0.5--1.5~keV (\xsoft), 0.5--4.5~keV (\xmed), and
1.5--6.0~keV (\xhard).}

$^{\rm d}${For sources CX1--CX9, \Lx\ comes from an average of
the unabsorbed luminosities of the best-fit models for each source
(see Tables~\ref{tab:xrayspec} and \ref{tab:cx1spec}).
A linear relation between \Lx\ and \xmed\ counts for these sources was
derived and used to estimate \Lx\ for sources CX10--CX19 based upon
their \xmed\ counts.  Typical uncertainties in \Lx\ are $\sim$20\%.}

$^{\rm e}${Type of counterpart (optical or radio) found and
associations (if any) with previously reported sources.  ``X'' numbers
refer to the \rosat\ sources of \citet{vj00}. ``B'' numbers refer to
the \hst\ sources of \citet{ba96}.  Note that our \hst\ positions for
these variables are very different from those given by Bailyn et al.\
(see \S\ref{sect:optastrom}).}

$^{\rm f}${See \S\ref{sect:res} for details.  ``Gal.'' indicates the
source may be a galaxy.}

\end{sidewaystable}

\begin{deluxetable}{lllll}
\tablewidth{0pt}
\tablecaption{Spectral fits to the \chandra\ data of the brighter
sources for fixed $\nh=\ee{2.2}{20}$~\pcmsq. \label{tab:xrayspec}}
\tablecolumns{5}
\tablehead{
\colhead{Src} & \colhead{Model\tablenotemark{a}} & \colhead{$kT$ (keV)} &
\colhead{\Lx\ (\ergsec)\tablenotemark{b}} & \colhead{$\chi^2_{\rm red.}$ (dof)} \\
\colhead{} & \colhead{} &\colhead{or $\Gamma$}& \colhead{[0.5--2.5~keV]}& \colhead{}
}
\startdata
CX1 & TB & 99.8\err{\infty}{66.4} & \ee{1.8}{32} & 1.04 (25) \\
    & BB & 0.66\err{0.05}{0.04} & \ee{2.0}{32} & 4.05 (25) \\
    & PL & 1.19\err{0.07}{0.09} & \ee{1.7}{32} & 1.17 (25) \\ \hline
CX2 & TB & 20.8\err{146}{72.4} & \ee{6.0}{31} & 1.02 (19) \\ 
    & BB & 0.45\err{0.08}{0.07} & \ee{6.0}{31} & 3.48 (19) \\ 
    & PL & 1.32\err{0.15}{0.13} & \ee{5.9}{31} & 1.03 (19) \\ \hline
CX3 & TB & 5.20\err{4.6}{1.9} & \ee{5.3}{31} & 1.20 (16) \\ 
    & BB & 0.50\err{0.08}{0.11} & \ee{5.5}{31} & 3.53 (16) \\ 
    & PL & 1.62\err{0.17}{0.17} & \ee{5.2}{31} & 1.23 (16) \\ \hline
CX4 & TB & 17.1\err{117}{10.2} & \ee{4.2}{31} & 1.05 (12) \\
    & BB & 0.61\err{0.10}{0.10} & \ee{3.6}{31} & 2.90 (12) \\
    & PL & 1.35\err{0.18}{0.17} & \ee{4.1}{31} & 1.14 (12) \\ \hline
CX5 & TB & $>$29 & \ee{3.5}{31} & 0.84 (14) \\
    & BB & 0.65\err{0.12}{0.11} & \ee{3.6}{31} & 1.07 (14) \\ 
    & PL & 0.93\err{0.25}{0.25} & \ee{3.5}{31} & 0.66 (14) \\ \hline
CX6 & TB & 11.7\err{126}{7.2} & \ee{2.2}{31} & 1.02 (10) \\ 
    & BB & 0.38\err{0.11}{0.07} & \ee{2.2}{31} & 2.41 (10) \\
    & PL & 1.41\err{0.24}{0.22} & \ee{2.2}{31} & 1.06 (10) \\ \hline
CX7 & TB & 34.2\err{\infty}{29.1} & \ee{2.0}{31} & 0.78 (8) \\ 
    & BB & 0.45\err{0.31}{0.13} & \ee{1.8}{31} & 3.12 (8) \\
    & PL & 1.33\err{0.35}{0.35} & \ee{1.9}{31} & 0.71 (8) \\ \hline
CX8 & TB & 6.32\err{\infty}{5.0} & \ee{2.2}{31} & 2.07 (6) \\
    & BB & 0.26\err{0.05}{0.04} & \ee{1.7}{31} & 1.35 (6) \\
    & PL & 1.31\err{0.48}{0.49} & \ee{2.3}{31} & 2.11 (6) \\ \hline
CX9 & TB & 3.34\err{115}{2.0} & \ee{1.3}{31} & 0.14 (4) \\
    & BB & 0.34\err{0.10}{0.07} & \ee{1.3}{31} & 0.49 (4) \\
    & PL & 1.64\err{0.48}{0.47} & \ee{1.3}{31} & 0.21 (4) \\
\enddata
\tablenotetext{a}{TB = Thermal Bremsstrahlung, BB = Blackbody, PL =
Power Law}
\tablenotetext{b}{Unabsorbed luminosity for $d=4.1$~kpc.}
\end{deluxetable}

\begin{deluxetable}{lllll}
\tablewidth{0pt}
\tablecaption{Spectral fits to the \chandra\ data of CX1 with \nh\
allowed to vary. \label{tab:cx1spec}}
\tablecolumns{5}
\tablehead{
\colhead{Model\tablenotemark{a}} & \colhead{\nh\ ($10^{20}$~\pcmsq)} & \colhead{$kT$ (keV)} &
\colhead{\Lx\ (\ergsec)\tablenotemark{b}} & \colhead{$\chi^2_{\rm red.}$ (dof)} \\
\colhead{} & \colhead{} &\colhead{or $\Gamma$}& \colhead{[0.5--2.5~keV]}& \colhead{}
}
\startdata
TB& 8.36\err{1.61}{3.21} & 15.9\err{21.0}{6.3} & \ee{2.1}{32} & 0.66 (24) \\
BB& $<0.48$ & 0.67\err{0.04}{0.04} & \ee{1.9}{32} & 3.70 (24) \\
PL& 10.4\err{4.0}{3.7} & 1.44\err{0.14}{0.13} & \ee{2.2}{32} & 0.73 (24) \\ \tableline
BB+PL& 1.99\err{10.9}{1.99}& \parbox{6.5em}{$kT=0.67\err{0.32}{0.17}$ \\ $\Gamma=1.23\err{0.36}{0.15}$}
& \ee{1.9}{32} & 0.54 (22) \\
\enddata
\tablenotetext{a}{TB = Thermal Bremsstrahlung, BB = Blackbody, PL =
Power Law}
\tablenotetext{b}{Unabsorbed luminosity for $d=4.1$~kpc.}
\end{deluxetable}

\begin{sidewaystable*}
\caption{Summary of color selection for optical/UV counterparts
\label{tab:res}} 
\begin{tabular}{ccccccccc} 
\tableline \tableline
Optical& \multicolumn{2}{c}{Offset from CXO
posn.\tablenotemark{a}} & $R_{675}$ &
$H\alpha_{656}-R_{675}$ & $B_{439}-R_{675}$ & $U_{336}$ &

$U_{336}-V_{555}$ & $nUV_{255}-U_{336}$\\
star ID \#      &in $\alpha$ (\arcsec) & in $\delta$ (\arcsec) &&&&&&\\
\tableline
1  &0.05        &0.05    & $19.36\pm0.02$ & $-0.24\pm0.03$ & $-0.96\pm0.02$ & $19.78\pm0.06$ & $ 0.24\pm0.08$ & $0.05\pm0.15$ \\
        &&&                                                     &                                               &                                               &                                       &$0.96\pm0.06$ &                \\
2 &--0.09       &0.07    & $19.55\pm0.01$ & $-0.86\pm0.07$ & $-0.23\pm0.02$ & $19.83\pm0.06$ & $0.40\pm0.06$ & $0.36\pm0.17$ \\
3 &0.06 &--0.11  & $21.48\pm0.03$ & $-0.07\pm0.08$ & $-0.19\pm0.05$ &$21.8\pm0.2$\tablenotemark{b}      & $0.2\pm0.2$\tablenotemark{b}  &       --      \\
4 &--0.01       &0.06    & $20.83\pm0.05$ & $-0.76\pm0.07$ & $0.25\pm0.13$ & $20.14\pm0.08$ & $-0.33\pm0.13$ & $-0.05\pm0.19$ \\
        &&&                                                     &                                               &                                               &                                       &$-0.75\pm0.09  $ &             \\
5 &--0.02       &0.00    & $19.13\pm0.01$ & $-0.12\pm0.03$ & $0.03\pm0.02$ & $19.72\pm0.06$ & $0.73\pm0.13$ & $1.82\pm0.19$ \\
        &&&                                                     &                                               &                                               &                                       &$0.68\pm0.24$   &              \\
6 &--0.09       &0.00    & $23.63\pm0.27$ & $-0.34\pm0.49$ & $-0.96\pm0.32$ &           --              &                       --                      &       --      \\
7 &0.01 &--0.06  & $21.44\pm0.06$ & $-1.40\pm0.08$ & $-0.10\pm0.10$ & $21.78\pm0.21$ & $ 0.08\pm0.26$ & --\\
        &&&                                                     &                                               &                                               &                                       &$0.48\pm0.21$ &                \\
10 &0.06        &--0.05  &--                                    &--                               &--                                   & $19.67\pm0.06$ &                & $0.97\pm0.19$ \\
        &&&                                                     &                                               &                                               &                                       &$-0.27\pm0.08$ &                                               \\
11 &--0.05      &0.05    & $21.60\pm0.28$ & $-0.41\pm0.35$ & $0.07\pm0.39$ &                    --              &                       --                      &       --      \\
13 &0.13 &0.16    & $24.26\pm0.15$ & $-0.06\pm0.47$ & $-0.70\pm0.20$ &           --              &                       --                      &       --      \\
15 &0.21        &--0.18  & --                                   &--                                     &--                             & $22.16\pm0.24$ & $ 0.16\pm0.25$ & $-0.62\pm0.54$ \\
16 &0.06 &--0.18  & $19.05\pm0.01$ & $-0.17\pm0.02$ & $-0.05\pm0.01$ &           --              &                       --                      &       --      \\
\tableline
\end{tabular}

$^{\rm a}${The optical position is then
given by \chandra\ X-ray value + offset in each coordinate.  Note, the
results are in true seconds of arc for both RA and Dec.}

$^{\rm b}${These magnitudes are estimates based on the wide U (F300W) and wide V (F606W) data available
($U_{300}=21.85\pm0.06$, $U_{300}-V_{606}=0.36\pm0.06$) and
transformations given in \citet{holt95}.}
\end{sidewaystable*} 

\begin{deluxetable}{lrcccccc}
\tablewidth{0pt}
\tablecaption{The Nature of the Faint Globular Cluster X-ray Sources.
\label{tab:compare}
}
\tablehead{
\colhead{Cluster} & \colhead{Coll.\ Freq.\tablenotemark{a}} & \colhead{qLMXB} &
\colhead{CV} & \colhead{Bin.\tablenotemark{b}} & \colhead{MSP\tablenotemark{c}} &
\colhead{Unc.\tablenotemark{d}} & \colhead{Ref.\tablenotemark{e}}
}
\startdata
47~Tuc  & $\equiv$100 & 2 & 13    & 5   & 9 & $\sim60$ &1 \\
$\omega$~Cen & 16      & 1 & 2     & 0   & 0 & $\sim 20$& 2 \\
NGC~6752 & 9           & 0 & 11-14 & 1-3 & 1 & 6 & 3 \\
NGC~6397 & 1           & 1 & 8     & 2   & 1 & 1 & 4\\
\enddata
\tablenotetext{a}{Collision frequency scaled on the frequency in
47~Tuc. It is computed based on the central
density $\rho_o$ and core radius $r_c$ given in the 1999 June 22
version of the catalogue described in \citet{har96},
as $\propto\rho_o^2r_c^3$.}
\tablenotetext{b}{X-ray-active main-sequence binaries.}
\tablenotetext{c}{X-ray detected MSPs.}
\tablenotetext{d}{Unclassified.}
\tablenotetext{e}{1.\ \citet{gr01a}; 2.\ \citet{rut01}; 3.\ this
paper; 4.\ \citet{gr01b}}
\end{deluxetable}

\end{document}